\documentclass[aps,prl,twocolumn,superscriptaddress,showpacs]{revtex4}
\usepackage{graphicx}

\begin{document}

\title{Spectral statistics in an open parametric billiard system}

\author{B. Dietz}
\affiliation{Institut f{\"u}r Kernphysik, Technische Universit{\"a}t
Darmstadt, Schlossgartenstr.~9, 64289 Darmstadt, Germany}
\author{A. Heine}
\affiliation{Institut f{\"u}r Kernphysik, Technische Universit{\"a}t
Darmstadt, Schlossgartenstr.~9, 64289 Darmstadt, Germany}
\author{A. Richter}
\altaffiliation{Electronic address: Richter@ikp.tu-darmstadt.de}
\affiliation{Institut f{\"u}r Kernphysik, Technische Universit{\"a}t
Darmstadt, Schlossgartenstr.~9, 64289 Darmstadt, Germany}
\author{O. Bohigas}
\affiliation{Laboratoire
de Physique Th{\'e}orique et Mod{\`e}les Statistiques, B{\^a}timent 100,
Universit{\'e} de Paris-Sud, 91405 Orsay Cedex, France}
\author{P. Leboeuf}
\affiliation{Laboratoire
de Physique Th{\'e}orique et Mod{\`e}les Statistiques, B{\^a}timent 100,
Universit{\'e} de Paris-Sud, 91405 Orsay Cedex, France}
\date{\today}

\begin{abstract}
We present experimental results on the eigenfrequency statistics of a
superconducting, chaotic microwave billiard containing a rotatable obstacle.
Deviations of the spectral fluctuations from predictions based on Gaussian
orthogonal ensembles of random matrices are found. They are explained by
treating the billiard as an open scattering system in which microwave power is
coupled in and out via antennas. To study the interaction of the quantum (or
wave) system with its environment a highly sensitive parametric correlator is
used.
\end{abstract}

\pacs{05.45.Mt, 03.65.Nk, 42.25.Bs}

\maketitle

Classical chaos manifests itself in universal spectral quantum fluctuations
that can be described by random matrix theory (RMT) \cite{bgs}.
While the earliest investigations of spectral correlations were confined to
nuclear physics \cite{nuclear}, during the last twenty years the universality
has been tested in other areas, like optical experiments \cite{doya}, quantum
dots \cite{marcus}, and acoustic setups \cite{ber99}. The (local) spectral
statistics depend generically only on the underlying symmetries of the system.
In particular, they are described by the Gaussian orthogonal
ensemble (GOE) of real symmetric random matrices for spinless systems with
time reversal symmetry, and by the Gaussian unitary ensemble (GUE) of complex
Hermitian random matrices in the absence of time reversal invariance \cite{note}.
The sensitivity of the quantum (or wave) statistical properties to fundamental
symmetries is obviously of great interest. For instance, it has been utilized
to derive an upper bound for the magnitude of the time or parity violating
component in nuclear interactions \cite{fkpt,losalamos}. 

We investigate here spectral properties of a superconducting microwave
resonator where currents are {\it induced by the measurement process}. 
Although we study a specific wave system, the results are expected to be
of general validity in the physics of complex quantum systems (atoms, molecules,
nuclei, quantum dots, ...).
The influence of the flux of microwave power flowing from the feeding to the
receiving antenna on the spectral properties of the system is so weak that it
can only be detected through a highly sensitive diagnosis tool, in our case a
parametric statistical measure. In a previous experiment, the wave system was
realized by a normal conducting microwave resonator attached to a large number
of antennas \cite{haake2}. There, the distribution of wave functions showed
significant deviations from the GOE predictions, which were attributed to the
transformation of the standing waves inside the closed microwave billiard into
waves propagating from an emitting antenna into a large number of exit
channels \cite{Saich}. The aim of the present paper is to go further into the
investigation of this mechanism. We will show that deviations from GOE
behaviour are already observed in a resonator with only three (or less)
attached antennas, when studying spectral properties as a function of a parameter.

\begin{figure}
\centerline{\includegraphics[width=4.5cm,height=4.cm]{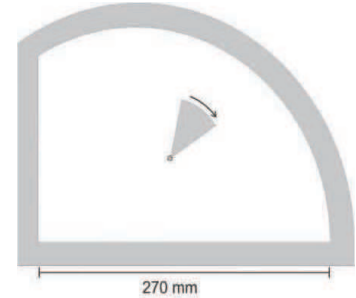}}
\caption{Sketch of the billiard, showing the outer boundary, 
as well as the rotatable
wedge-shaped piece of teflon. The angle of rotation $\alpha$ of the wedge
defines the parameter (its initial orientation is arbitrary).}
\label{resonator}
\end{figure}

The experiment discussed here has been performed with a superconducting
microwave resonator, whose high-quality factor is typically $Q=10^{5}$ or larger
\cite{imapro}, i.e. dissipative processes in the resonator are reduced to a
minimum, thereby ensuring a high spectral resolution. Results obtained with a
flat cylindric resonator are presented. Aside their intrinsic interest, such
resonators mimic two-dimensional quantum billiards of corresponding shapes
\cite{imapro,stoeckmannbuch,sto90sri91gra92,nobel}. The analogy is based on
the isomorphism between the scalar Helmholtz equation of the electric field
$\vec{E}$ for wavelengths longer than twice the height of the resonator, where
$\vec{E}(\vec{r})=\Psi(x,y)\vec{e}_{z}$ is perpendicular to the billiard
($xy$) plane, and the Schr{\"o}dinger equation for the wavefunctions in the
quantum billiard. The eigenvalues $k_i^2$ of a closed resonator satisfy the
Helmholtz equation with Dirichlet boundary conditions imposed on $\Psi(x,y)$.
They are directly related to the eigenenergies of the corresponding quantum
billiard. In this analogy, the Poynting vector plays the role of the quantum
probability current density \cite{vbvrs}.

The microwave resonator has been manufactured from lead-plated copper, as in
\cite{dreieck}; its shape is shown in Fig.~\ref{resonator}. During the
measurements it has been placed in a liquid helium cryostat at a temperature
of $T=4.2$~K, which guarantees superconductivity of the lead surface. The
outer boundary of the resonator has the shape of a desymmetrized straight-cut
circle. The dynamics inside the corresponding classical billiard is chaotic
\cite{reichl}. A dielectric wedge of teflon inside the resonator has been
rotated with a leverage from outside the cryostat. The cavity has been coupled
to one feeding and two receiving antennas.
They are tiny metal pins of 0.5~mm in diameter and 
have been adjusted such that they mechanically reach only some 100
microns into the interior of the cavity guaranteeing weak coupling and hence
minimal disturbance of the excited field in the resonator. Using a HP-8510C
network analyzer the transmission spectra have been measured in the frequency
range up to 18~GHz for 37 equidistant settings of the angle $\alpha$ (cf
Fig.~\ref{resonator}) in steps of 2.5 degrees.

\begin{figure}
\centerline{\includegraphics[width=5.5cm,height=4.5cm]{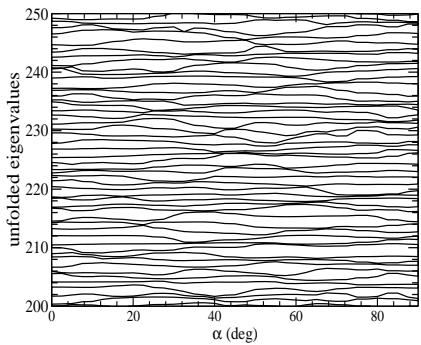}}
\caption{Part ($\# ~200-~\#~250$) of the experimental eigenvalues on an
unfolded, i.e. dimensionless scale as a function of the parameter $\alpha$.}
\label{ewdyn}
\end{figure}

Complete eigenvalue sequences of 440 resonances for each value of the
parameter $\alpha$ have been measured. Figure \ref{ewdyn} shows the
eigenvalues number 200 to 250 as a function of the parameter, the so-called
eigenvalue dynamics or parametric fluctuations, where for each of the 37
spectra the eigenvalues have been unfolded by scaling them to unit mean
spacing \cite{imapro,stoeckmannbuch}. As the parameter $\alpha$ is varied, the
positions of the resonance frequencies (i.e. the real part of the resonance)
describe irregular oscillatory curves that generically do not cross each
other. The oscillations have a mechanical interpretation, namely the
derivative $-\partial k_i^2/\partial\alpha$ is proportional to the torque
exerted by the i-th electromagnetic eigenmode on the teflon wedge.
For the statistical analysis of the spectra the whole set of resonance
frequencies has also been unfolded with respect 
with respect to the parameter $\alpha$ by following 
the proceedure described in \cite{ls}.
This allows to properly incorporate the characteristic scales associated with
the frequency and parameter secular variations \cite{ls}, thereby defining
dimensionless quantities.

\begin{figure} 
\centerline{\includegraphics[width=5.5cm,height=4.5cm]{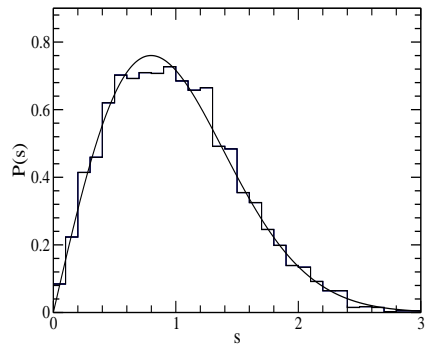}}
\caption{The experimental nearest-neighbor spacing distribution (histogram);
the GOE random matrix result (full line). The eigenvalues $\#~331$~to~$\#~440$ 
of each of the 37 spectra have been used.}
\label{p_s_exp}
\end{figure}

In Fig.~\ref{p_s_exp} the spacing distribution $P(s)$ of the distance between
consecutive eigenvalues computed at fixed values of $\alpha$ is shown. The
experimental $P(s)$ is in good agreement with the GOE result. A similar
agreement with GOE is found for the number variance $\Sigma^2$ and the least
mean square statistics $\Delta_3$. This is in agreement with common
expectations, especially in the present experiment with a high $Q$ value of
the resonator and a small coupling to the antennas. However, a minimum
coupling is unavoidable (open system), thereby implying a presumably small
perturbation of the closed system. The purpose of the present investigation is
the identification of signatures of this disturbance produced by the
measurement process in the spectral properties.

It is by now well established that in systems depending on parameters,
correlations between eigenvalues at different parameter values lead to
important extensions of RMT universalities \cite{alt,Pato}. In the present
experiment, these correspond to correlations between spectra at different
orientations of the teflon wedge. Several parametric correlators
\cite{Alhassid} have been computed, namely the velocity distribution, the
curvature distribution, the velocity--velocity correlator, the diffusion
correlator and the distribution of the spacings at avoided crossings (a
general presentation of the results will be given elsewhere). We focus here on
the latter, i.e. the probability distribution $P(c)$ of the local minima $c$
of the distance between neighboring levels as the parameter $\alpha$ is
varied.

The parameter dependence of the chaotic resonator is modeled by the following
ensemble,
\begin{equation} \label{hmu}
\hat H(\mu)=\cos\mu\cdot\hat H_0+\sin\mu\cdot\hat H_1 \ ,
\end{equation}
where $\hat H_0,\ \hat H_1$ are $N$-dimensional GOE random matrices and $\mu$
is a real parameter. Before performing statistical analyses, the spectra are
unfolded (with respect to energy and parameter dependence, see above). We will
denote by $\mu_{resc}$ the resulting rescaled parameter. Though the form for
large $N$ of the probability distribution $P(c)$ has not been derived for this
model, it is well approximated by the $N=2$ result \cite{zakrzewski91}
\begin{equation} \label{p_c_goe} 
P_{\rm \scriptscriptstyle GOE}(c) \approx \frac{2}{\pi}\exp(-\frac{c^{2}}{\pi})
\ ,
\end{equation}
where the scale of $c$ has been chosen such that $\langle c \rangle=1$.

\begin{figure} 
\centerline{\includegraphics[width=5.5cm,height=4.5cm]{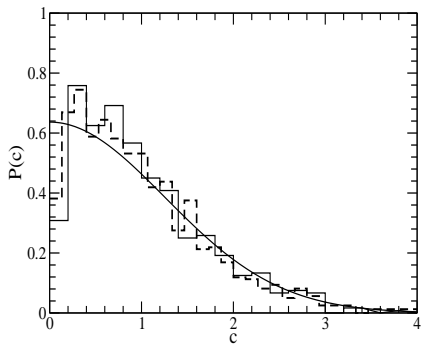}}
\caption{The experimental distribution of avoided crossings (full line
histogram) compared to the GOE prediction Eq.(\ref{p_c_goe}) (full line
curve). The dashed line histogram is obtained from the model (\ref{heff}) with
$\lambda =0.02$ and using a discrete set of rescaled parameters $\mu_{resc}$ 
(see main text). The eigenvalues $\#~331$~to~$\#~440$ of each of the 37 spectra 
have been used.}
\label{minabst}
\end{figure}

We have evaluated $P(c)$ from the experimental data. The result is presented
in Fig.~\ref{minabst}. It is compared to the prediction Eq.~(\ref{p_c_goe}).
Though the general trend shows an overall agreement, systematic deviations are
observed. In particular, a lack of small spacings of avoided crossings is
clearly visible in the experimental distribution. 

It is interesting to note that similar results were obtained in
\cite{stoeckmann97}, where they were attributed to an insufficient
experimental resolution in the parameter variation. This was also our first
suspicion and we tested it numerically with a random matrix model. The
rotatable obstacle, i.e. the varying of the shape of the resonator, is
accounted for in the model by considering the parameter dependent ensemble of
random matrices given in Eq. (\ref{hmu}), where the dimension of the matrices
$\hat H_0,\ \hat H_1$ is chosen as $N=1000$. Only the 300 central eigenvalues
of each diagonalization are used. If the rescaled parameter $\mu_{resc}$ is
treated as a continuous parameter then a curve for $P(c)$ very similar to
Eq.(\ref{p_c_goe}) results. If, in contrast, $\mu_{resc}$ is discretized such
that a set of parameter values very similar to the experimental one is used,
then the agreement with the experimental histogram improves. In particular, a
dip at $c \sim 0$ appears. However, systematic deviations of the experimental
$P(c)$ and also of other parametric statistical measures studied (velocity and
curvature distribution, velocity-velocity correlator) from the corresponding
predictions of the model (\ref{hmu}) persist. We thus conclude that the
disagreement between the RMT predictions of the model (\ref{hmu}) and the
experimental results cannot be entirely attributed to an insufficient
resolution with respect to the parameter. Other effects, like the stability of the 
dip with increasing frequency, were also checked. We have devided the 400 levels 
in windows of increasing frequency and computed $P(c)$ for each of them. We were 
not able to detect any systematic
trend in the behavior of the dip as frequency increases. The present available
data thus exclude the attribution of the dip to a finite size effect.

Since the experiment is performed with a superconducting resonator, in
the absence of the antennas and neglecting dissipation, the cavity is an
isolated time--reversal invariant system which, in principle, should be
correctly described by the parametric statistical model Eq. (\ref{hmu}).
However, for the measurement of a spectrum, a typical procedure is to couple
the system to the exterior through antennas, and to emit an input signal via one of
them and receive the output signal via another one. Hence, the effective
Hamiltonian describing the spectral properties of the resonator is
non-Hermitian (open system), and the statistical properties of the spectrum
are not expected to coincide with those predicted from the Hermitian model Eq.
(\ref{hmu}). If the coupling to the antennas is weak, a small but nonzero
change in the position of the real part of the resonances with respect to the
closed system is thus expected. As Fig.~\ref{p_s_exp} shows, with the present
experimental conditions this shift in the resonance frequencies has no visible
effect on the spectral fluctuations \cite{strgth} for a fixed value of the
parameter. In contrast, as we will see below, it induces sizeable deviations
in the parametric statistical properties.

In the present experiment  
the antennas act as single scattering channels as their diameter is small
compared to the wavelengths of the microwaves in the total frequency range.
Wave scattering in such a three-port system is described by a $3 \times 3$
scattering matrix of the form \cite{Albeverio,Haake,Seba,Ott}
\begin{equation}
\hat S=\hat I+2i\hat W^T (\hat H(\mu)-i\hat W\hat W^T -E\hat I)^{-1}
\hat W \ ,
\label{Smatrix}
\end{equation} 
whose derivation is based on the theory of quantum scattering
(formulated e.g. in \cite{Mahaux}).
Here, $\hat I$ is the identity matrix and $\hat H(\mu)$ the Hamiltonian of the
resonator. It is modeled by the $N \times N$ parametric GOE defined in
Eq.~(\ref{hmu}). The matrix $\hat W$ is an $N \times 3$ matrix, $\hat W
\propto (\hat X_1,\hat X_2, \hat X_3)$, that describes the coupling of the resonator 
to
the antennas (the $j$-th component of the $N$-dimensional column vector
$\hat X_\sigma$ couples the $j$-th internal wavefunction to the $\sigma$-th
antenna). From Eq.(\ref{Smatrix}), the resonances are obtained as the
eigenvalues of the effective non-Hermitian Hamiltonian \cite{note2}
\begin{equation} \label{heff}
\hat H_{eff}(\lambda,\mu)= \hat H (\mu) - i\lambda{\pi\over\sqrt{N}}\left(\hat
X_1\hat X_1^T+\hat X_2\hat X_2^T+\hat X_3\hat X_3^T\right) .
\end{equation}
Since the system is time reversal symmetric and the coupling is weak, the
emission of waves from one antenna and its detection in another is modeled
with real column vectors $\hat X_\sigma$ as in \cite{Albeverio}. Consistently
with the random model adopted for the Hamiltonian of the resonator, they are
considered as independent random variables with a Gaussian distribution whose
width is set to unity. Then, in Eq. (\ref{heff}) the parameter $\lambda$
measures the strength of the coupling of the resonator to the antennas in
units of the mean spacing $\pi/\sqrt{N}$ of the eigenvalues of $\hat H(\mu)$.
In contrast to the present experiment, the strong coupling regime has been 
investigated in \cite{Haake,Lenz}.

We have studied numerically the statistical properties of the eigenvalues of
$\hat H_{eff}$. For small values of $\lambda$ the resonances are close to the
real axis and tend to the eigenvalues of $\hat H(\mu)$. As $\lambda$
increases, and up to $\lambda \sim 0.5$, the imaginary part of the resonances
increases. For larger values, the resonances split into two groups: three of
the resonances, their number corresponding to the rank of the perturbation, move 
deeply into the complex plane while the remaining $N-3$ approach again
the real axis with increasing $\lambda$. Due to the weak coupling, the present
experiment should correspond to relatively small values of $\lambda$. We find
numerically that for values of $\lambda$ smaller than $\approx 0.05$ the ratio
of the imaginary to the real part of the eigenvalues of $\hat H_{eff}$ is
smaller than 0.005, and its ratio to the mean spacing between adjacent real 
parts is less than 0.1. The numerical
$P(s)$, the $\Delta_3$-statistics and the $\Sigma^2$-statistics agree with
GOE, in accordance with the experimental results. The distribution $P(c)$ of
avoided crossings, however, deviates from the model (\ref{hmu}) for $\lambda$
larger than about $0.01$. Before reaching again a GOE--like behavior
(\ref{p_c_goe}) at $\lambda\simeq 2.5$ as tested numerically, a sharp
$\delta$-like peak at $c\simeq 0$ followed by a dip is observed. The peak size
increases until $\lambda \simeq 0.5$, and then decreases. This behavior differs
both from the GOE behavior, Eq.(\ref{p_c_goe}), and from the experimentally
observed distributions. A more careful analysis of the behaviour of the real
and the imaginary part of the eigenvalues of $\hat H_{eff}$, Eq.~(\ref{heff}),
as a function of the parameter $\mu $ shows that the contributions to the peak
of $P(c)$ at $c=0$ are due to crossings of the real parts of two complex
eigenvalues. Incidentally, this behaviour is characteristic of non-Hermitian
Hamiltonians studied recently \cite{EP}.

Why is the sharp peak at $c \sim 0$ predicted by the model (\ref{Smatrix}) not
observed in the experiment? Our interpretation is that this is due to the
discrete variation of the experimental parameter. Because of the discrete
sampling, the probability to observe small spacings is strongly reduced. And
indeed, the smallest spacing observed in the unfolded experimental data is
about $s\simeq 0.014$. To compare theory and experiment a discretization of
the parameter in the model (\ref{heff}) has to be performed. In
Fig.~\ref{minabst} such a comparison is made using in Eq.(\ref{heff}) a
discrete set of parameter values $\mu_{resc}$ whose step size is similar to
the experimental one, and $\lambda = 0.02$. A good overall agreement between both
curves is obtained, not only around the dip close to the origin. A similar
agreement is obtained for the other parametric correlators studied (curvature
distribution, velocity-velocity correlator, etc), thus providing a globally
consistent picture of the experimental data.

The experimental as well as the theoretical results thus indicate that, while
absent in the nearest neighbor spacing distribution, we are observing in more
sensitive parametric spectral functions signatures induced by the measurement
process. Interestingly, similar mechanisms were recently studied in the physics
of cold atoms \cite{yoo}. Two ingredients are important in order to understand
the experimental results. First, we had to model the system by incorporating its
coupling to the external world (flux is fed into the resonator and coupled out
via one or two antennas). Second, we had to take into account the discreteness of the
parameter variations. 

The proposed parametric "spectral detector'' is clearly a powerful tool to
study the interaction of a quantum (or wave) system with its environment. The
underlying working principle is very general since it depends on a fundamental
physical principle, namely the standing waves inside the closed resonator are
transformed into waves propagating from an entrance antenna to an exit
antenna. In particular, it may be useful in the analysis of more
controlled experiments concerning the interplay between the measurement
process, the currents it induces through the cavity, and the dissipative
processes.

We thank C.~Dembowski and H.-D.~Gr{\"a}f for their ideas and suggestions
concerning the experiment. Discussions with T.~Guhr, U.~Kuhl, M.~P.~Pato,
H.-J.~St\"ockmann and H.~A.~Weidenm\"uller are acknowledged. We are grateful
for the kind hospitality and financial support of the 
Institut f\"ur Kernphysik, Darmstadt, and the Laboratoire de Physique 
Th\'eorique et Mod\`eles Statistiques, Orsay, which is an
Unit\'e de Recherche associ\'ee au CNRS, during several visits. 
This work was supported by the DFG within the SFB 634.


\begin{thebibliography}{99}
\bibitem{note2} The principal integral appearing in the effective Hamiltonian
Eq. (4.4.20b) of \cite{Mahaux} vanishes in a scattering system consisting of a 
microwave resonator and antennas, because there $\hat W\hat W^T\propto 1/k$ 
with $k$ denoting the wave number \cite{Albeverio}.
\end{thebibliography}
\end{document}